\documentclass[prl,twocolumn,amsfonts,showpacs]{revtex4-1}

\pdfoutput=1
\usepackage{epsfig}
\usepackage{psfrag}
\usepackage{amsmath}
\usepackage{amssymb}
\usepackage{color}
\usepackage{appendix}
\usepackage{bibunits}
\usepackage{hyperref}

\begin{document}
\author{Chuan-Yin Xia$^{1,2}$}
\author{Hua-Bi Zeng$^{1}$ \\ email: \href{mailto:hbzeng@yzu.edu.cn}{hbzeng@yzu.edu.cn}}

\affiliation{$^1$ Center for Gravitation and Cosmology, College of Physical Science
and Technology, Yangzhou University, Yangzhou 225009, China}
\affiliation{$^2$ School of Science, Kunming University of Science and Technology, Kunming 650500, China}


\title{Winding up a finite size holographic superconducting ring beyond Kibble-Zurek mechanism}
\begin{abstract}
We studied the dynamics of the order parameter and the  winding numbers $W$ formation of a quenched  normal-to-superconductor state
phase transition in
a finite size holographic superconducting ring. There is a critical circumference $\tilde{C}$ below it  no winding number will be formed,
then $\tilde{C}$ can be treated as the Kibble-Zurek mechanism (KZM)  correlation length $\xi$ which is proportional to the
fourth root of its quench rate $\tau_Q$,
which is also the average size of independent pieces formed after a quench.
When the  circumference $C \geq 10 \xi$, the key KZM scaling between  the average value of absolute winding number
and the quench rate $\langle|W|\rangle \propto \tau_Q^{-1/8}$ is observed. At smaller sizes,
the universal scaling will be modified, there are  two regions.
The middle size $5\xi<C<10\xi$ result $\langle|W|\rangle \propto \tau_Q^{-1/5}$ agrees with
a finite size experiment observation. While at $\xi<C\leq 5\xi$ the the average value of absolute winding number equals to
the variance of winding number and there is no well exponential relationship between the quench rate and the average value of absolute winding number. The winding number statistics can be derived  from a trinomial distribution with $\tilde{N}=C/ (f \xi)$ trials, $f\simeq 5$
is the average number of adjacent pieces that are effectively correlated.
\end{abstract}
\maketitle
\begin{bibunit}

Second order phase transition that traverses the critical point at a finite rate is a non-equilibrium process.
Due to the relaxation time's divergence near  critical point (¡®¡®critical slowing down¡¯¡¯),
the formation of topological defects by taking into account finite speed of propagation of the relevant information was predicted by the Kibble-Zurek mechanism (KZM)\cite{1,2,3,4,5,6,7,8}.
The original idea of KZM is from  Kibble's  insight on the role
of causality in structure formation in the early universe \cite{1,2}.
Later, Zurek found  that condensed-matter
systems offer a test-bed to study the dynamics of symmetry breaking \cite{3,4,5}. He firstly  predicted
the formation of independent regions with their average size controlled by the
correlation length $\xi$ at the point when the frozen time ends, $\xi$ scales with
the linear quench rate $\tau_Q$ in which the phase transition is crossed
as a universal power-law
\begin{equation}
\xi \propto \tau_Q ^{\alpha},
\label{eq1}
\end{equation}
 the power-law exponent
$\alpha =  \nu/(1 + \nu z)$ is set by a combination of  the dynamic and correlation-length
(equilibrium) critical exponents denoted by $z$ and $\nu$, respectively.
By assuming the number of defects is proportion to the numbers of independent
regions formed $N$ and $N \propto L^d/ \xi^d$ ($L$ is the size of the system, $d$ is the dimensionality of the system ), then
the average density $\tilde{n}=N/L^d$ of the resulting topological defects scales with
the linear quench rate
as a universal power-law
\begin{equation}
 {\tilde n} \propto \tau_Q^ {-d \alpha},
 \label{eq2}
\end{equation}
 this is the key prediction of KZM.

In 3D and 2D systems, the topological defects are usually vortex strings \cite{Antunes} and vortices \cite{Hindmarsh,Yates} respectively with
zeroth order parameter inside the vortex cores.
In a 1D system with real
order parameter, kinks is the topological defects with zero order parameter in the center \cite{Laguna}.
While in a 1D systems with a complex order parameter undergo a phase transition that breaks
the $U(1)$ gauge symmetry, winding numbers $W=\oint_C d\theta /2 \pi$ are expected to form but the amplitude of the order
parameter keeps uniform\cite{Das,Sonner:2014tca}.
Numerical experiments in a quenched superfluid/superconductor have supported
the KZM's key prediction of scaling in 3D  \cite{Antunes}, 2D \cite{Yates,Chesler:2014gya,Zeng} and 1 D \cite{Sonner:2014tca,Das}.
In laboratory, the Eq.(\ref{eq2}) has also been confirmed in liquid crystals \cite{Chuang,Bowick,Digal}, $^3$He superfluids \cite{Baeuerle,Ruutu}, Josephson junctions
\cite{Carmi,Monaco1,Monaco2,Monaco3}, thin-film superconductors \cite{Maniv,Golubchik},
a linear optical quantum
simulator\cite{Xu} and also in a  strongly interacting
Fermi superfluid \cite{Ko}.
Recently, Adolfo del Campo, etl found the universal  statistics of topological defects formed in a quantum phase transition beyond KZM \cite{Campo,Fernando}, which has been confirmed by a 1D quantum simulation \cite{Cui}.
\begin{figure}[t]
\centering
\includegraphics[trim=5.5cm 10.4cm 5.9cm 11.9cm, clip=true, scale=0.9, angle=0]{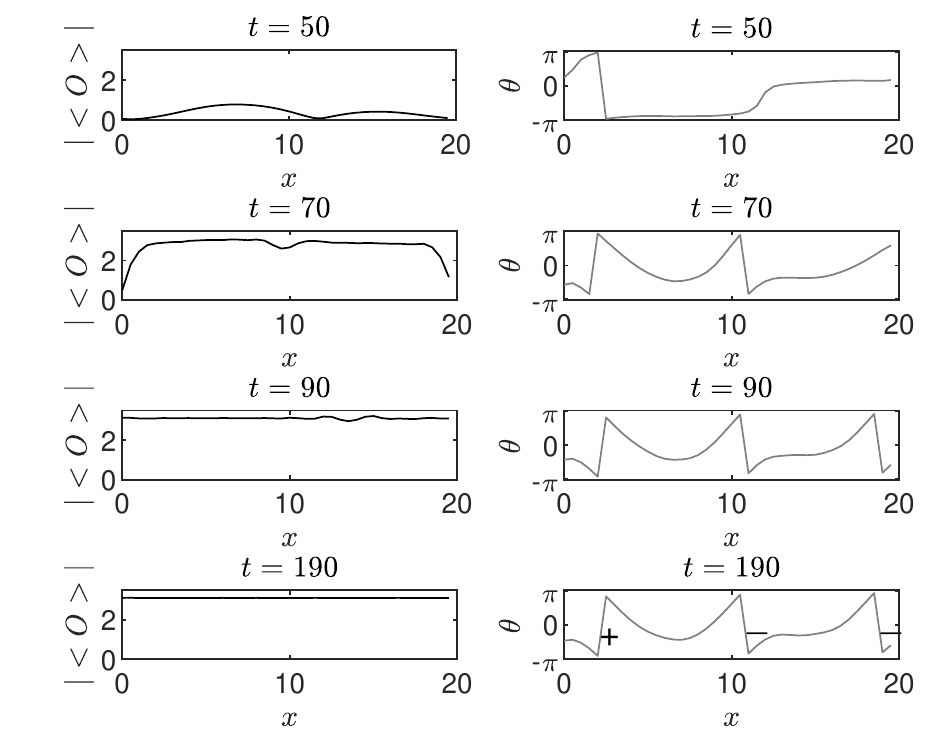}
\caption{\textbf{Winding up a superconducting ring after a temperature quench \textbf{$\tau_Q =e^3$}, the circumference $C=20$.}
In the four rows we show the
magnitude of the order parameter $|\langle O \rangle|$ and it's phase $\theta$ configuration in the dynamic process.
The system finally enters an equilibrium state with constant amplitude of order parameter and a fixed configuration
of phase field $\theta(x)$.
The winding number $W=n^+-n^-=-1$, $n^+=1,n^-=2$. }\label{fig1}
\end{figure}

Despite the progress, few attention has been devoted to the case when the size of the system
approaches the order of $\xi$. Except in experiment \cite{Corman} the size effect results was reported
in  an Bose gases through a temperature quench
of the normal-to-superfluid phase transition.  Large size observation matches the KZM's
prediction by using the mean field theory critical exponents $z=2, \nu=1/2$, where $\langle |W|\rangle \propto \tau_q^{-1/8}$ at fixed $C$ and $\langle |W|\rangle \propto \sqrt{C}$ at a fixed $\tau_Q$.
However, at small size $C < 10 \xi$, $\langle |W|\rangle \propto \tau_Q^{-0.2}$ at fixed $C$ and $\langle |W|\rangle \propto C^{0.8}$ at a fixed $\tau_Q$.
Now a theoretical model and even a numerical simulation is still lacking to address the experimental observation.
Also, the KZM does not include the case when the system size is not an integer multiple of $\xi$ since at large
sizes the remainder can be ignored, the number of KZM pieces $N$ is always an integral. However, at a finite size close to $\xi$, one have to consider the case of a fractional $N$.
Furthermore, equilibrium phase transitions at finite size can still have  universal scaling laws, as confirmed in
a $^3$He superfluid phase transition \cite{Francis} and a strongly coupled holographic superconductor \cite{Antonio},
then it is  of  importance to study the finite size effect on dynamics and defects formation in  phase transitions
crossed at a finite rate.

In order to study the finite size Kibble-Zurek mechanism, we adopt a holographic superconductor ring model in the framework of Gauge/Gravity duality, focusing on the spontaneous formation of winding in the  superconducting ring with any value of circumference after a temperature quench. The Gauge/Gravity duality\cite{Maldacena,Witten,Gubser} that relates
strongly interacting quantum field  theories to theories of classic gravity
in higher dimensions has been proved to be a new and useful scheme  to study
 strongly interacting condensed matter systems in equilibrium\cite{Zaanen:2015oix,Ammon2015},
and also to study the real time dynamics when the system is far away from equilibrium \cite{Liu:2018crr,Liu,Tian}.
Then it is very suitable to study the phase transition dynamics happened at a finite rate\cite{Sonner:2014tca,Bhaseen,Chesler:2014gya,Zeng,Das1,Natsuume,Basu1,Basu2,Murata}.
We adopt the well studied holographic superconductor model defined in a AdS black hole \cite{Gubser2008,Hartnoll,Herzog},
\begin{equation}
S=\int  d^4x \sqrt{-g}\Big[-\frac{1}{4}F^2-( |D\Psi|^2-m^2|\Psi|^2)\Big].
\end{equation}
This is the  Einstein-Maxwell-complex scalar model, where the black hole background geometry is $ds^2=\frac{\ell^2}{z^2}\left(-f(z)dt^2-2dtdz + dx^2+ dy^2\right)$ in the
Eddington coordinate, $f(z)=1-(z/z_h)^3$, the black hole temperature $T=3/(4 \pi z_h)$. $x,y$ are the boundary spatial coordinates.
There is a critical value of the black hole temperature below that the charged
scalar develops a finite value in the bulk while it's dual field theory operator have a finite expectation value
$\langle O \rangle$, which breaks the $U(1)$ symmetry in the boundary field theory.
Working in 1D spatial boundary geometry by only turning on coordinate $x$ dependence of all the
fields in the equations of motion, using the periodic boundary condition  in the $x$ coordinate we are effectively studying a superconducting
ring.
By solving the dynamic equations by changing the black hole temperature
above $T_c$ to a one below $T_c$, the quench induced winding number formation process
can be monitored in details\cite{append1}. One
sample result of a quench from $1.1T_c$ to $0.82 T_c$  is given in Fig.\ref{fig1}, where $\tau_q=e^3$ and $C=20$.
After the quench the order parameter amplitude approaches the equilibrium value while the
phase develops a stable configuration that does not change anymore, the winding number $W=-1$. From the phase configuration in Fig.\ref{fig1}, we can define the number of local positive winding numbers $n^+$ and
also  the number of local negative winding numbers $n^-$ as shown in the last plot, then
$W=n^+ - n^-=1-2=-1$. The $n^+$ and $n^-$ are important to understand the statistics of $W$ as will be
shown later.

The  winding number $W$ of a long quenched superconducting ring admits a Gaussian distribution then it's average value is always zero \cite{Das}.
In order to see how quench rate affect the formation of $W$, people always refer to it's variance $\sigma^2(W)$, which is
proportional to the number of independent pieces $N = C/\xi \propto \tau_q^{-1/4}$ using the mean field exponents $z=2$ and $ \nu=1/2$
in Eq.(\ref{eq1}), $\xi$ is the correlation length corresponding to the
time when the frozen time end. At large $N$ limit, the mean absolute winding number $\langle |W| \rangle$  can be computed from the Gaussian distribution, then one can obtain the key prediction \cite{Das}
\begin{equation}
 \sigma(W)=\sqrt{\langle W \rangle^2} \propto \langle |W| \rangle \propto \tau_Q^{-1/8} .
 \end{equation}
At a fixed rate, one can have $\langle |W| \rangle \propto {\sigma(W)} \propto \sqrt{N}$.

\begin{figure}[t]
\centering
\includegraphics[trim=2.3cm 11.0cm 3.8cm 9.7cm, clip=true, scale=0.6, angle=0]{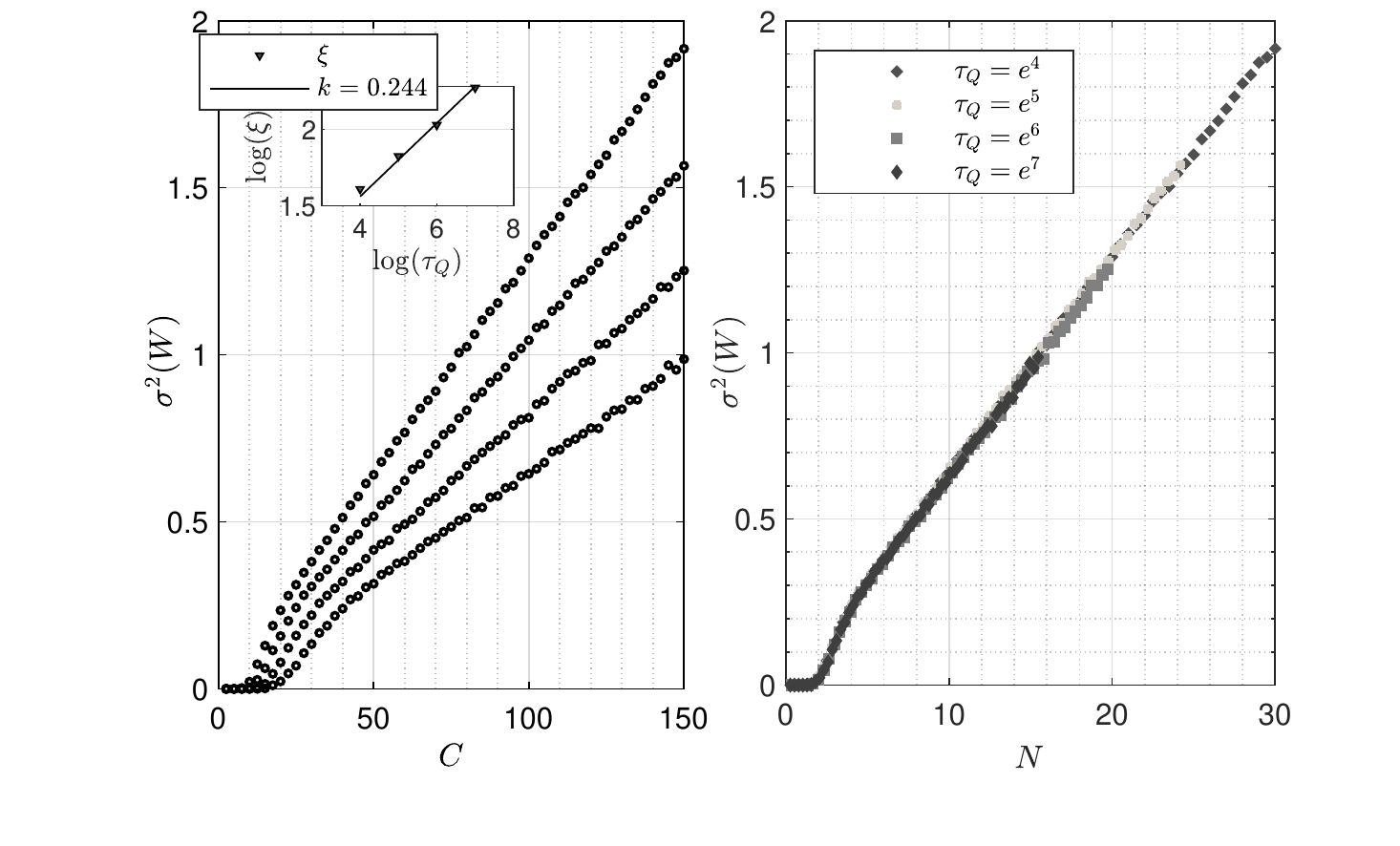}
\caption{ \textbf{Size dependence of  variance $\sigma^2(W)$}. (left)
Size dependence of $\sigma^2(W)$ for four quench rates, from top to bottom, $\tau_Q= e^4, e^5,e^6,e^7$.
The inset shows the scaling of the critical circumference. (right) $\sigma^2(W)$ as a function of  pieces number $N$,
all quench rate results are identical to each other.}\label{fig2}
\end{figure}

Fig.\ref{fig2} (left) plots $\sigma^2(W)$  as a function of $C$ for four different quench rates ($\tau_Q= e^4, e^5,e^6,e^7$),
average over $100000$ times calculation for a circumference.
According to KZM, since the system inherits an infinite relaxation time
at the critical point then it can not catch the speed of a quench,
as a result, the superconducting ring will be divided into
many independent pieces with their average size to be $\xi(\tau_Q)$.
The topological defects forms at the positions where the independent pieces meet,
it is nature to conclude that there will be a  critical value of circumference $\tilde{C}$ that below which there is
only one piece then no winding number will be formed. $\tilde{C}$ is the average size of an
independent region, also the value of the correlation length $\xi(\tau_Q)$.
Numerical result of $\tilde{C}$ confirmed the key prediction of KZM that  $\xi(\tau_Q) \propto \tau_Q^{1/4}$ (Fig.\ref{fig2} inset).
The Log-Log plot  founds that the slop $k=0.224$, close to $0.25$ as expected.
According to KZM's picture of the quenched dynamic phase transition,  $\sigma^2(W)[C]$ lines for different rates can be scaled to exactly one
line by transforming  $C$ to be the number of regions/pieces formed (Fig.\ref{fig2} (right)), in a formula it can be expressed as
\begin{equation}
 \sigma^2(W_{\tau_{Q1}})[C_1]=\sigma^2(W_{\tau_{Q2}})[C_2],
\end{equation}
when $C_1/\xi(\tau_{Q1})=C_2/\xi(\tau_{Q2})=N$, $N$ is the number of independent pieces formed after the quench.
Also the linear relationship $\sigma^2(W) \propto N$ between variance of winding number and number of regions
can be seen when the size of regions is roughly larger than five.
Then, we confirmed the KZM mechanism from another perspective from the size dependent results rather than
computing Eq.(\ref{eq2}) directly. Note that the agreement also confirmed the  fact that a  holographic superconducting phase transition
is always of the mean field class \cite{Maeda,Jensen,Zeng2018}.

\begin{figure}[t]
\centering
\includegraphics[trim=3.3cm 10.0cm 0.0cm 9.7cm, clip=true, scale=0.6, angle=0]{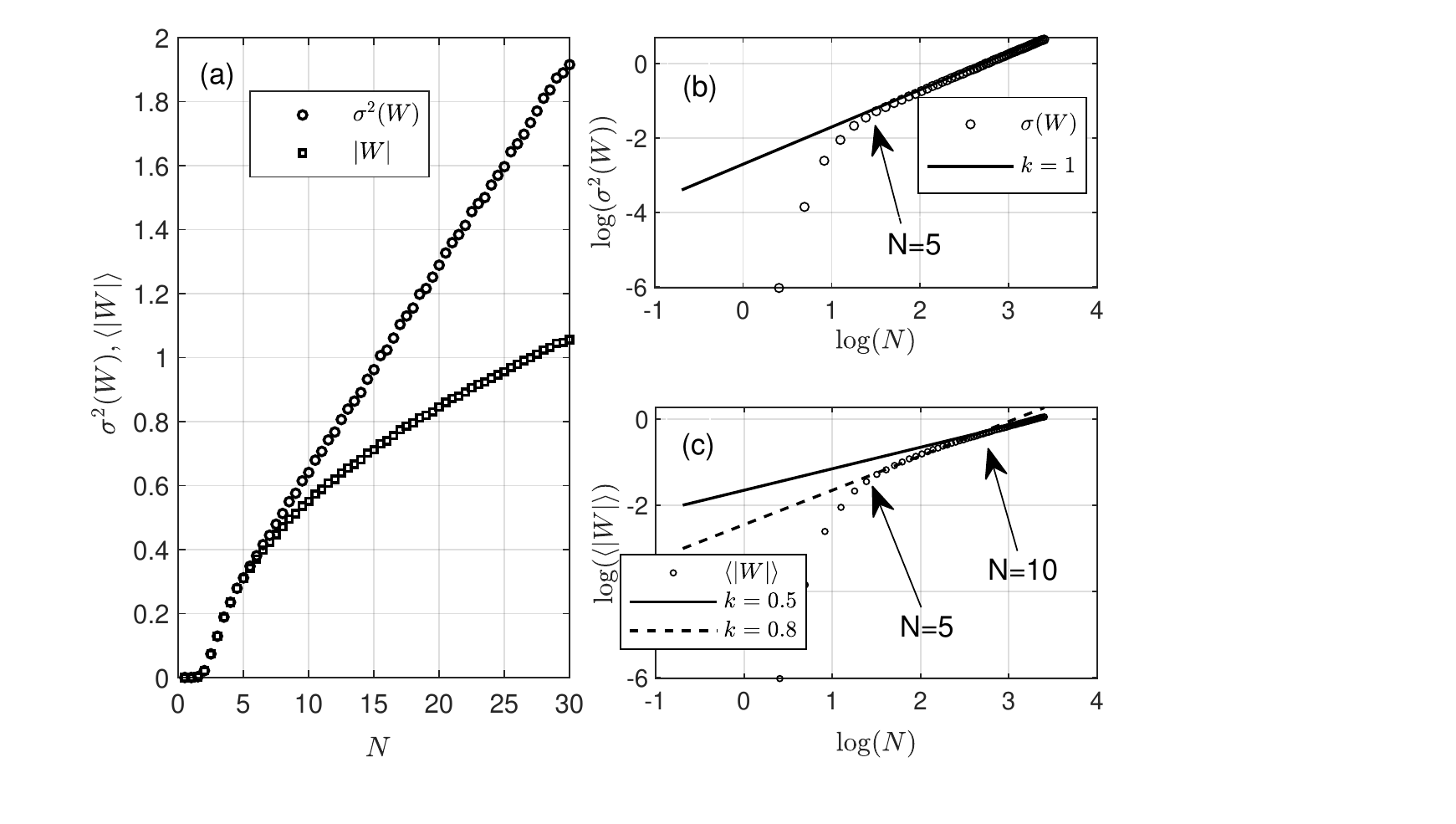}
\caption{\textbf{Scaling of $\sigma^2(W)$ and $\langle |W| \rangle$}. (a) $\sigma^2(W)$ and $\langle |W| \rangle$
as a function of $N$, when $N<5$ the two have the same values. (b) Logarithmic relationship between  $\sigma^2(W)$ and $N$. (c) Logarithmic relationship between  $\langle |W| \rangle$
and $N$. $k$ is the linear fit slop. }\label{fig3}
\end{figure}

Besides the perfectly KZM matched results, from Fig.(\ref{fig2})
one can  also find that the prediction $\sigma^2(W) \propto {N}$ can not hold when the ring size is reducing.
From Fig.(\ref{fig3}) (b) the log-log plot tells that the non KZM region is $ 1< N \leq 5$.
Also in this region, there is another interesting feature that $\sigma^2(W)=\langle |W| \rangle$,
since the winding number can have only three value $-1,0,1$.
with the vanishing $\langle W \rangle$. One thing needs to emphasize is that in this region,
though the KZM pieces $N$ can be larger than one, $W$ does not take a value larger than one, this may
indicates that the formation of topological defects in a size $N \leq 5$ is effectively correlated.

\begin{figure}[t]
\centering
\includegraphics[trim=5.3cm 9.0cm 0.0cm 9.5cm, clip=true, scale=0.7, angle=0]{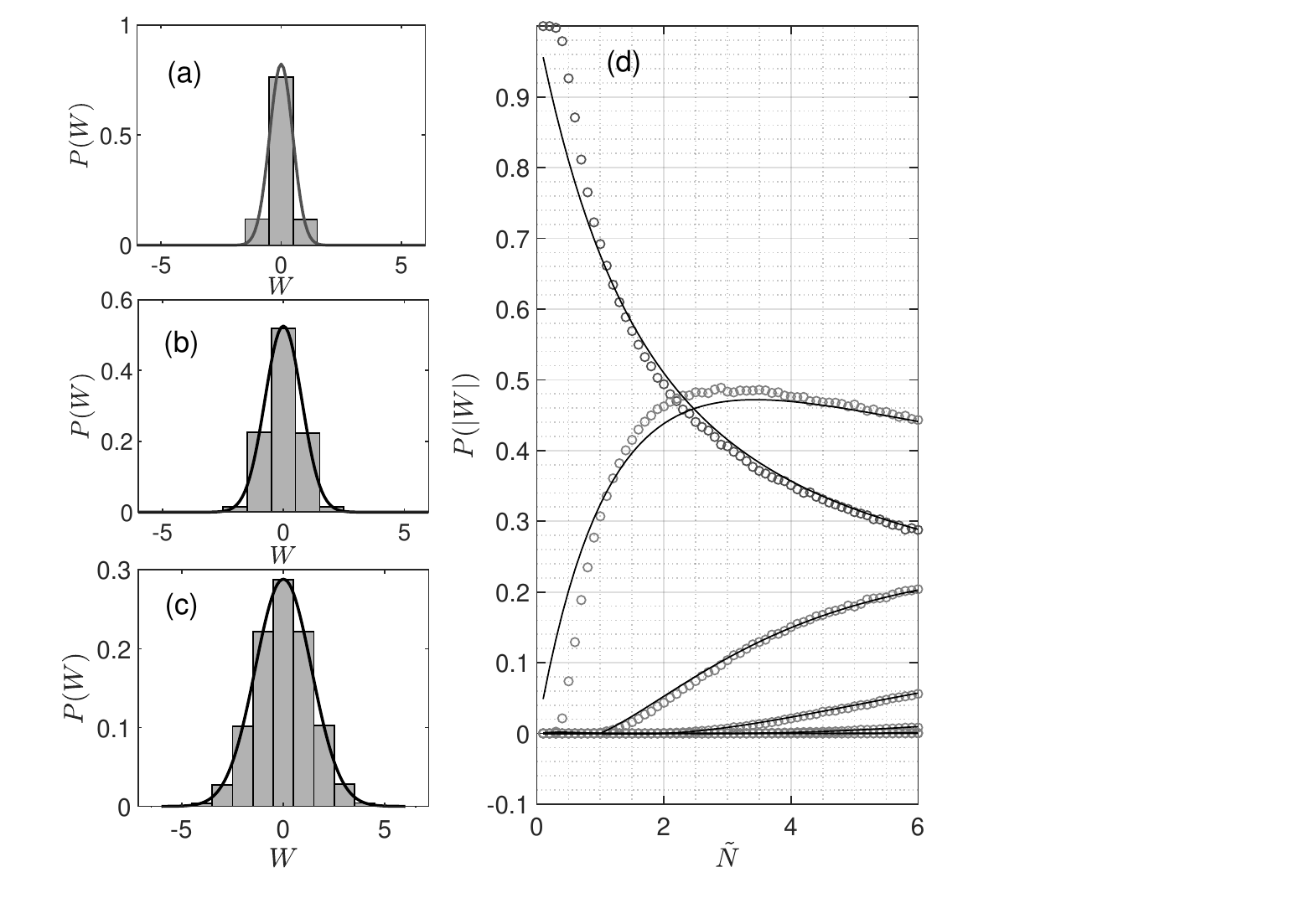}
\caption{\textbf{$P(\langle |W| \rangle)$ from trinomial distribution.}(a-c)Histogram of $P(W)$ for $N=8,18,30$ respectively when $\tau_Q=e^4$. (d) Numerical results of $P(|W|)$ (circles, from top to bottom are $|W|=0,1,2,3,4,5,6$ respectively) and their fitted curves from Eq.8 (Solid lines).}\label{fig4}
\end{figure}

Also from Fig.\ref{fig3} (b,c), at a larger size $5<N<10$, $\sigma^2(W)$
has a linear dependence  of $N$, but
\begin{equation}
\langle |W| \rangle \propto N^{0.8}.
\label{eq6}
\end{equation}
This is different from the large size result $\langle |W| \rangle \propto N^{0.5}$, which is
a result that the $W$ admit a Gaussian distribution \cite{Das}.
Then one can conclude that  the Gaussian
distribution is not  good any more to capture the statistic distribution of winding number at the small size case.
We compared the discrete distribution and the corresponding Gaussian distribution with the
variance of $W$ in Fig.(\ref{fig4}), which shows
the  winding distribution at  three different numbers of regions $N=8,18,30$.
The discrete distribution approaching a Gaussian one when
increasing $N$. While $N<10$, $|W|$ has only three values from zero to two,
the deviation from Gaussian distribution is expected.
Combining Eq. \ref{eq6} and
\begin{equation}
\xi=\frac{C}{N}\propto \tau_Q^{1/4},
\end{equation}
we get exactly the scaling between average absolute winding number and quench rate beyond KZM reported in experiment \cite{Fernando} when $5<N<10$
\begin{equation}
\langle |W| \rangle \propto \tau_Q^{-0.2}.
\end{equation}

Due to the fact that the global winding number $W=n^+-n^-$, the winding number
in local regions can be  $+$, $-$ or $0$, it is nature to
expect the distribution of local winding number $n^+,n^-$ can be captured
by a trinomial distribution. The probability of the trinomial distribution of trials $\tilde{N}$ reads
\begin{equation}
P(\tilde{N},n^+,n^-)=\frac{\tilde{N}!}{n^0 ! n^+! n^- !}(\frac{p}{2})^{n^++n^-}(1-p)^{n^0},
\label{eq9}
\end{equation}
where $n^0=\tilde{N}-(n^+ + n^-), n^0$ is the number of pieces where there is no local winding number.
$(n^+,n^-,n^0) \leq \tilde{N}$, $\tilde{N}$ equals to largest number of $n^+(n^-)$, which can be
defined as the number of effectively unrelated pieces. $p/2$ is the probability for both $``+"$ and $``-"$ since the two have the same
distribution \cite{append2}, while $1-p$
is the $``0"$ probability.
From the largest number of $n^+(n^-)$ of a fixed $N$ we find that $\tilde{N}=n^+_{max}=N/5$.
Furthermore, we consider the case when the $\tilde{N}$ is not a integral,
by increasing from $\tilde{N}$ from a smaller integral number $M$ to $M+1$,
the $\sigma^2(W)$ is increasing continuously without a jump.
To understand the continuous of $\sigma^2(W)$,
we apply the mathematical theorem that the factorial in Eq.(\ref{eq9}) can be expressed by the
Gamma function
\begin{equation}
M!=\Gamma(M+1)=\int_0^\infty y^M e^{-y} dy.
\end{equation}
To understand the fractional number $N$ of KZM pieces formed after a quench,
we conject that the number of pieces formed  admit a normal distribution with it's
average value equal to $N$, which can be a any value fractional number.
With the distribution, one can calculate the probability $P(|W|)$, for example, $P(W=0)$ is the summation
of the cases where $n^+=0, n^-=0$ and $n^+=n^-$ in Eq.(\ref{eq9}). $P(|W|=1)$ equals to the summation
of the cases where $n^+=n^- \pm 1$.
$P(|W|)$ as a function of $\tilde{N}=N/5$ obtained from the trinomial distribution can basically  match the numerical results as shown in Fig. \ref{fig4} (d), the best fit parameters were found to be $p=0.324$ and $\tilde{N}=N/5$,
due to the fact that when $N=30$, $|W|$ has it's maximal value $|W|_{max}=6$.
The $P(|W|=0)$ will keep decreasing when $|W|$ takes  larger maximal values due to the increasing numbers of pieces.
$P(|W|=1)=0$ when $N<1$, when $N>1$ $P(|W|=1)$ keeps increasing to its maximal value at about $\tilde{N}=2$.
$P(|W|=2)=0$ when $N<5(\tilde{N}<1)$. In a word, $P(|W|=M)$ will have a finite value only when $N>5M (\tilde{N}>M)$.
Another check of the trinomial distribution can be done by computing $\sigma^2(W)$ and $\langle |W| \rangle$
from $P(W)$ obtained from Eq.(\ref{eq9}) \cite{append3},
compare the results to the numerical simulation in Fig. \ref{fig3}
we find good agreement when $\tilde {N} \geq 2$.
The deviation is obvious when $\tilde{N}<1$, in this region with only one effectively independent KZM piece,
there are only three possibilities: $n^+=n^-=0; n^+=1, n^-=0$; and $n^+=0, n^-=1$.
Then $\langle |W| \rangle= \sigma^2 (W)= P(n^+=1)+ P(n^-=1)= p$, the probability $p$ is increasing from
zero  at $N=1$ to a constant $p=0.324$ at $N=5$.


In summary, our numerical experiment on spontaneous formation of winding numbers in
a finite size ring not only confirms the KZM predictions but also presents new   findings of dynamics of phase transition
in a superconducting ring. Firstly, the finite size  distribution of $W$  is different from Gaussian distribution then the   KZM scaling law
of $\langle |W| \rangle$ will be modified, which agrees the experimental observation \cite{Corman}. Secondly,
the KZM assumes the pieces formed after a quench are independent, while the numerical results
at finite size indicates that the five adjacent pieces are correlated, then the effective  independent pieces $\tilde{N}= L/5 \xi$.
Furthermore, a continuous version of trinomial distribution with a fractional  $\tilde {N}$ trial was proposed to  understand
the size dependent statistic distribution of winding numbers. The nonlinear dependence between $\sigma^2(W)$ and $N$ is believed
to be a results that the winding number formation process are dependent when $N \leq 5$, where the probability $p$ equals to $\sigma^2(W)$ which is
 $C$ dependent. The number of effectively independent  pieces is $\tilde{N}=N/5$, where $\tilde{N}$ still admits the scaling law predicted by  Kibbe-Zurek mechanism.

\emph{Acknowledgements}.---
We thank Wei Can Yang for valuable discussions.
This work is supported by the National Natural
Science Foundation of China (under Grant No. 11675140).

\end{bibunit}

~
\newpage

\begin{appendices}

\begin{bibunit}

\section{SUPPLEMENTAL MATERIAL}

\subsection{Equation of motion  and the numerical scheme to simulate a temperature quench}
With the ansatz  $ \Psi = \Psi(t, z, x), A_t = A_t(t, z, x), A_x = A_x(t, z, x),
A_y = A_z = 0$, and define $\Phi=\Psi/z$.
The implicit form of the Euler-Lagrange equations from Einstein-Maxwell-complex scalar model reads
\begin{widetext}
\begin{eqnarray}
\partial_t \partial_z \Phi - i A_t \partial_z \Phi - \frac12 [ i \partial_z A_t \Phi + f \partial_z^2 \Phi + f' \partial_z \Phi - z \Phi &&
\nonumber\\
+ (\partial_x^2 \Phi ) - i \partial_x A_x  \Phi - A_x^2 \Phi - 2 i (A_x \partial_x \Phi) ] &=& 0;
\\
\partial_t \partial_z A_t - (\partial_x^2 A_t + \partial_y^2 A_t) - f \partial_z (\partial_x A_x + \partial_y A_y) + \partial_t (\partial_x A_x + \partial_y A_y) &&
\nonumber\\
+ 2 A_t |\Phi|^2 - i f (\Phi^* \partial_z \Phi - \Phi \partial_z \Phi^*) + i (\Phi^* \partial_t \Phi - \Phi \partial_t \Phi^*) &=& 0;
\\
\partial_t \partial_z A_x - \frac12 \left[ \partial_z (\partial_x A_t + f \partial_z A_x) - i (\Phi^* \partial_x \Phi - \Phi \partial_x \Phi^*) - 2 A_x |\Phi|^2 \right] &=& 0;
\end{eqnarray}
\end{widetext}
There is another constrain equation from the time component of Maxwell equations,
\begin{equation}
\partial_z (\partial_x A_x - \partial_z A_t) + i (\Phi^* \partial_z \Phi - \Phi \partial_z \Phi^*) = 0.
\end{equation}
Throughout this paper, we work in the units with $e = c = \hbar = k_b = 1$ and we also have scaled $\ell=1$.

There is a critical black hole temperature below which the solution with a finite $\Phi$
then the $U(1)$ symmetry can be spontaneous broken, all the critical exponents
at the phase transition point are of the standard mean field values.

To solve the highly non-linear PDEs, the boundary condition for the charged scalar and the gauge field
must be imposed, specifically, at the infinite boundary,
\begin{eqnarray}
A_x(t,z,x)&=& a_x(t,x)+ b_x(t,x) z+\mathcal{O}(z^2),
\label{aboundary} \\
 \Psi(t,z,x)&=& \Psi_1(t,x) z+ \Psi_2(t,x) z^2+\mathcal{O}(z^3).
 \label{psiboundary}
\end{eqnarray}
from the AdS/CFT correspondence dictionary the coefficients $a_{x}$ can be regarded as the gauge field on the boundary along $x$ directions while $J_{x}=-b_x-(\partial_x a_t-\partial_t a_x)$ is the  supercurrent.
 By fixing $J_{x}$ we are dealing with a superconducting ring with dynamic gauge field $a_{x}$ \cite{Skenderis,Witten,Domenech, Montull}. Coefficients $a_t$ and $b_t$ are interpreted as chemical potential $\mu$ and charge density $\rho$  respectively in the boundary field theory. Moreover, $\Psi_1$ is a source term which is set to be zero, then $\Psi_2$ is the vacuum expectation value $\langle O \rangle$ of the dual scalar operator in the boundary in the spontaneous symmetry broken phase. The temperature of the black hole is $T=3/(4\pi z_h)$,
technically, to tune the temperature people usually set $z_h=1$ while changing the value of $\rho$ according
a scale symmetry of the equation. The dimensionless quantity relate temperature and the $\rho$ is
$T/\rho^2$, there is critical $\rho_c\approx 4.06$ above which the system will enter a lower free energy state with non-vanishing $\Psi$, quenching a $\rho$ across $\rho_c$ is equal to quench the system from a high temperature state to a low temperature state in a linear way by the following
setup of a time dependent $\rho(t)$ in the quenching stage before approaching the final temperature $T_f < T_c$
\begin{equation}
\rho(t)=\rho_c(1-\frac{t}{\tau_Q})^{-2}.
\end{equation}
$t=0$ is defined as the moment quench begins, before quenching the superconductor,
the system with initial fluctuations had been thermalized for a sufficient time.
The initial random seeds of the fields in the bulk by satisfying
the statistical distributions $\langle s(t, x_i)\rangle = 0$ and $\langle s(t, x_i)s(t', x_j)\rangle = h \delta(t-t')\delta(x_i-x_j)$, with the amplitude $h = 10^{-3}$.  In the radial direction z, we use the Chebyshev pseudo-spectral method with 21
grids. Since in the $x$-directions, all the fields are periodic, we use the Fourier decomposition
along $x$-directions. Filtering of the high momentum modes are implemented following the
``2/3's rule'' that the uppermost one third Fourier modes are removed \cite{chesler1}.

\begin{figure}[t]
\centering
\includegraphics[trim=2.3cm 6.0cm 0.0cm 6.7cm, clip=true, scale=0.4, angle=0]{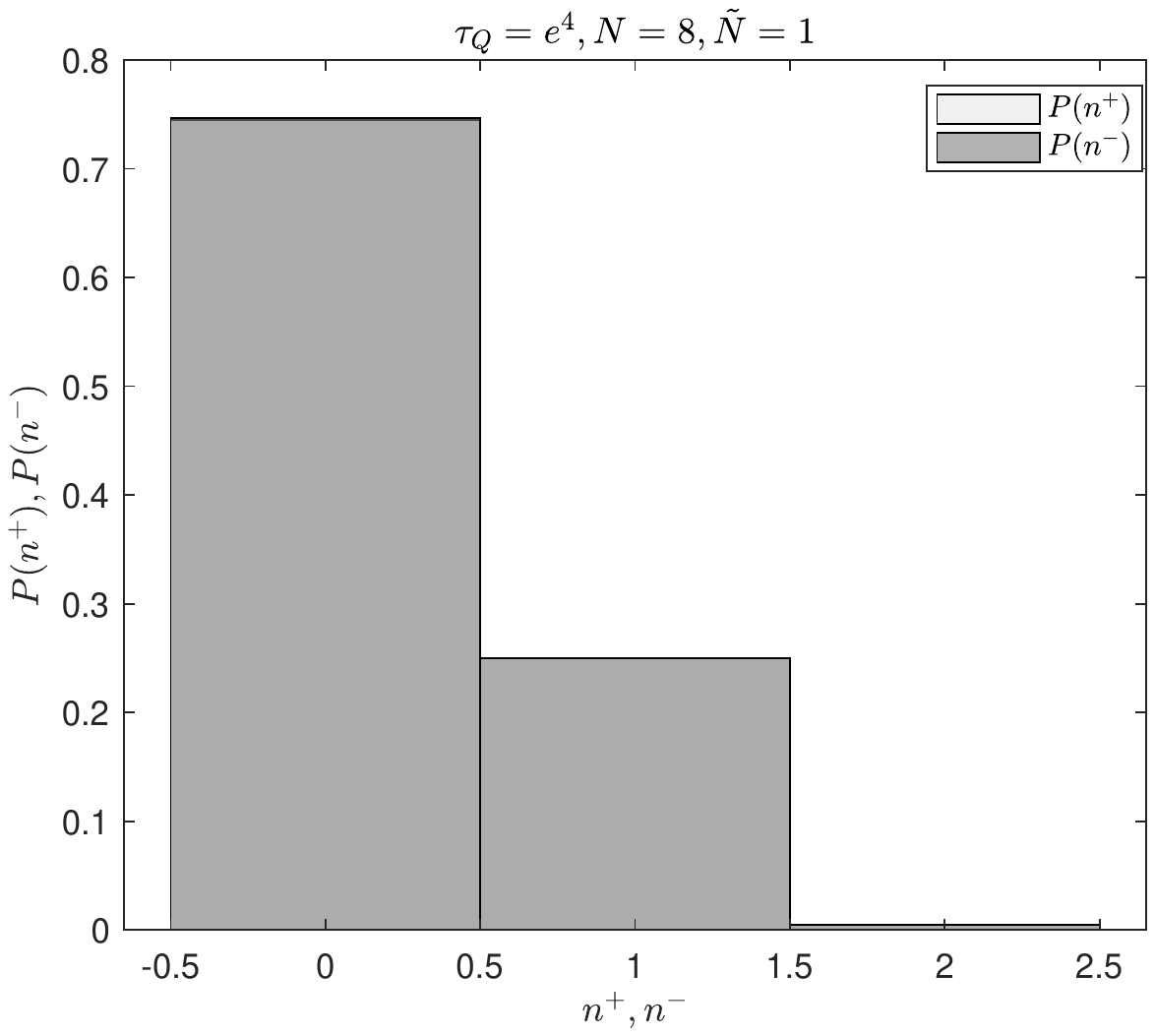}
\includegraphics[trim=2.3cm 6.0cm 0.0cm 6.7cm, clip=true, scale=0.4, angle=0]{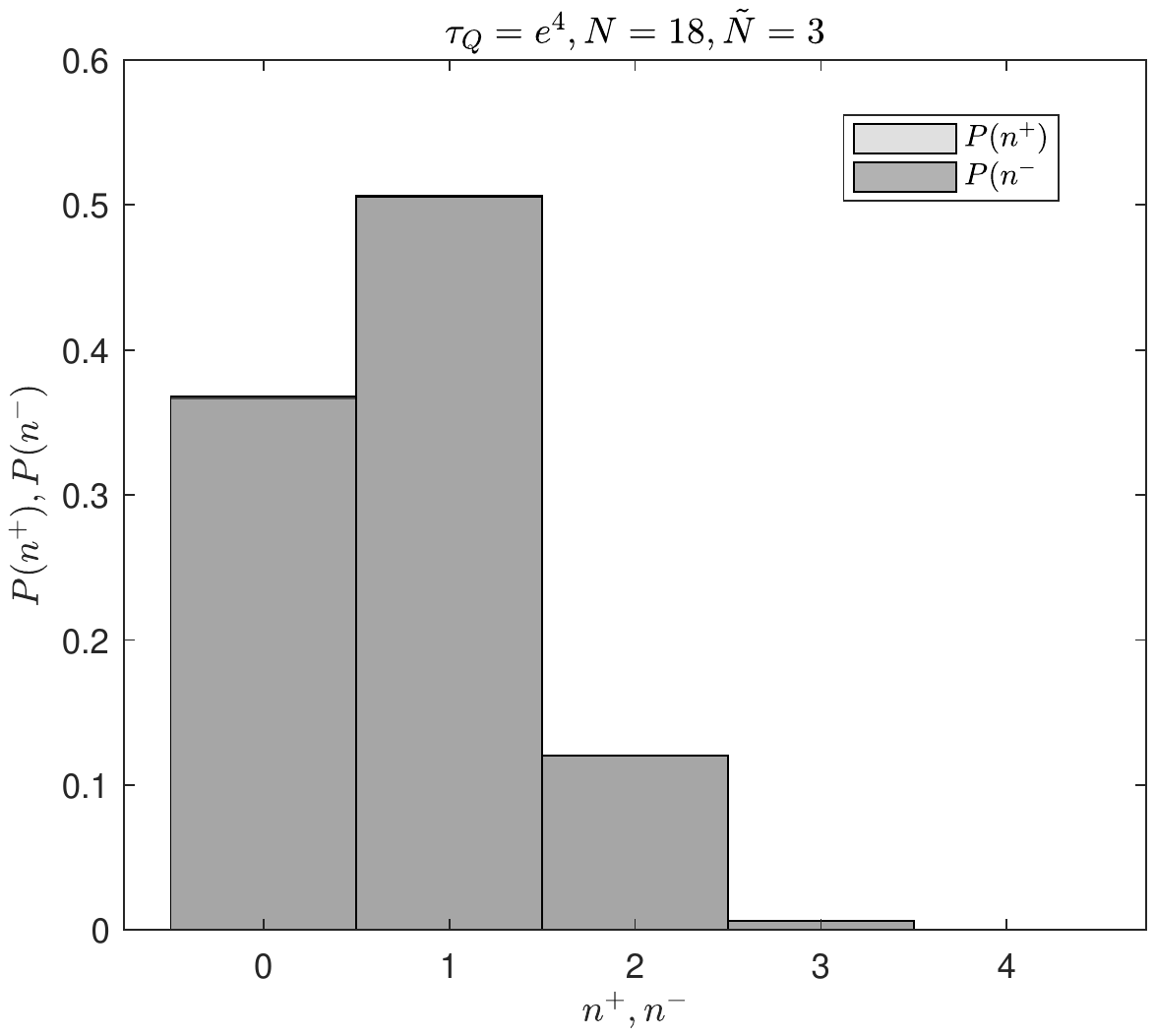}
\includegraphics[trim=2.3cm 6.0cm 0.0cm 6.7cm, clip=true, scale=0.4, angle=0]{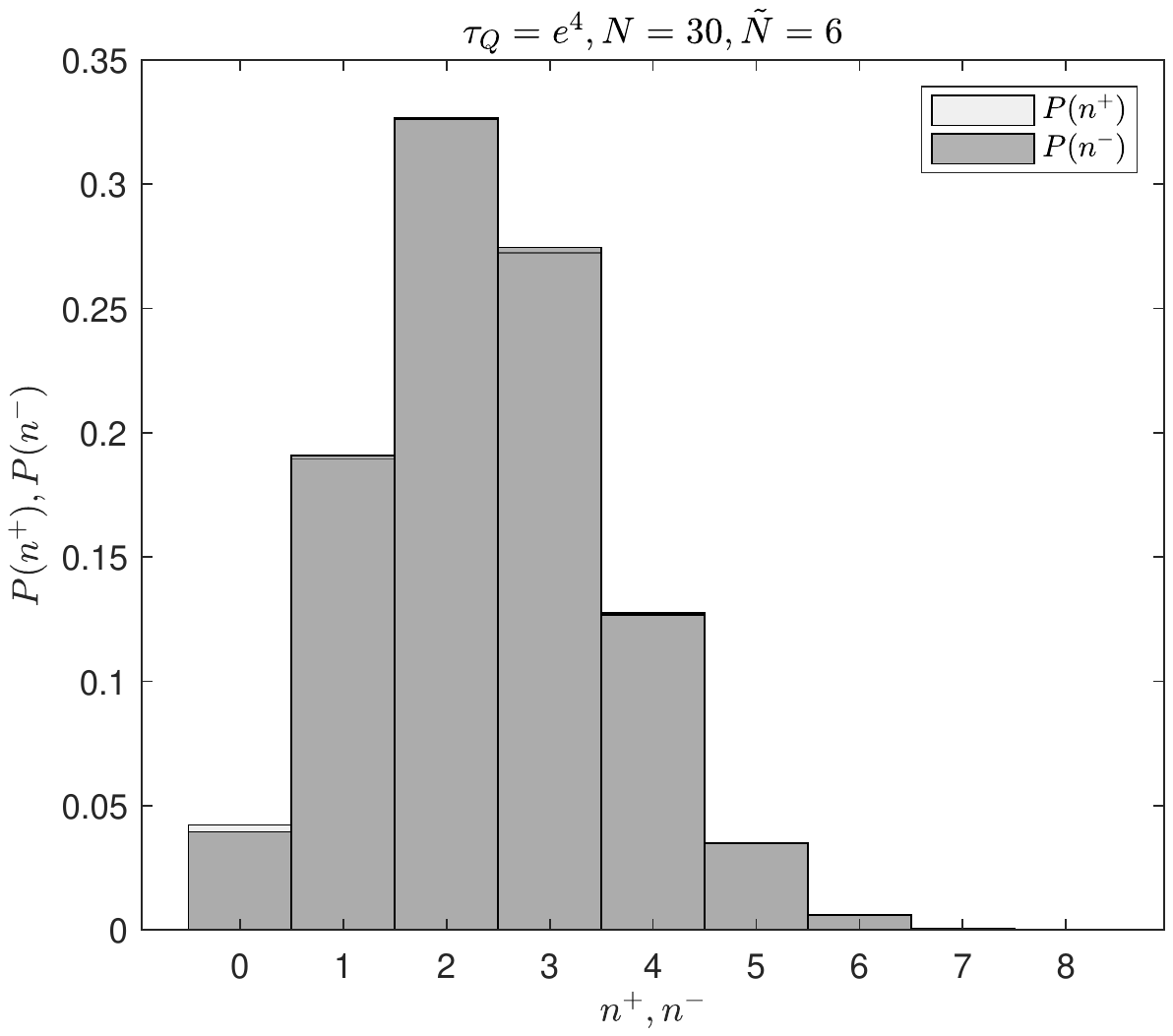}
\caption{\textbf{Distribution of $n^+$ and $n^-$ for three different sizes.} $P(n^+)$ always equals to $P(n^-)$.}\label{fig5}
\end{figure}

\begin{figure}[t]
\centering
\includegraphics[trim=2.3cm 11.0cm 0.0cm 6.7cm, clip=true, scale=0.8, angle=0]{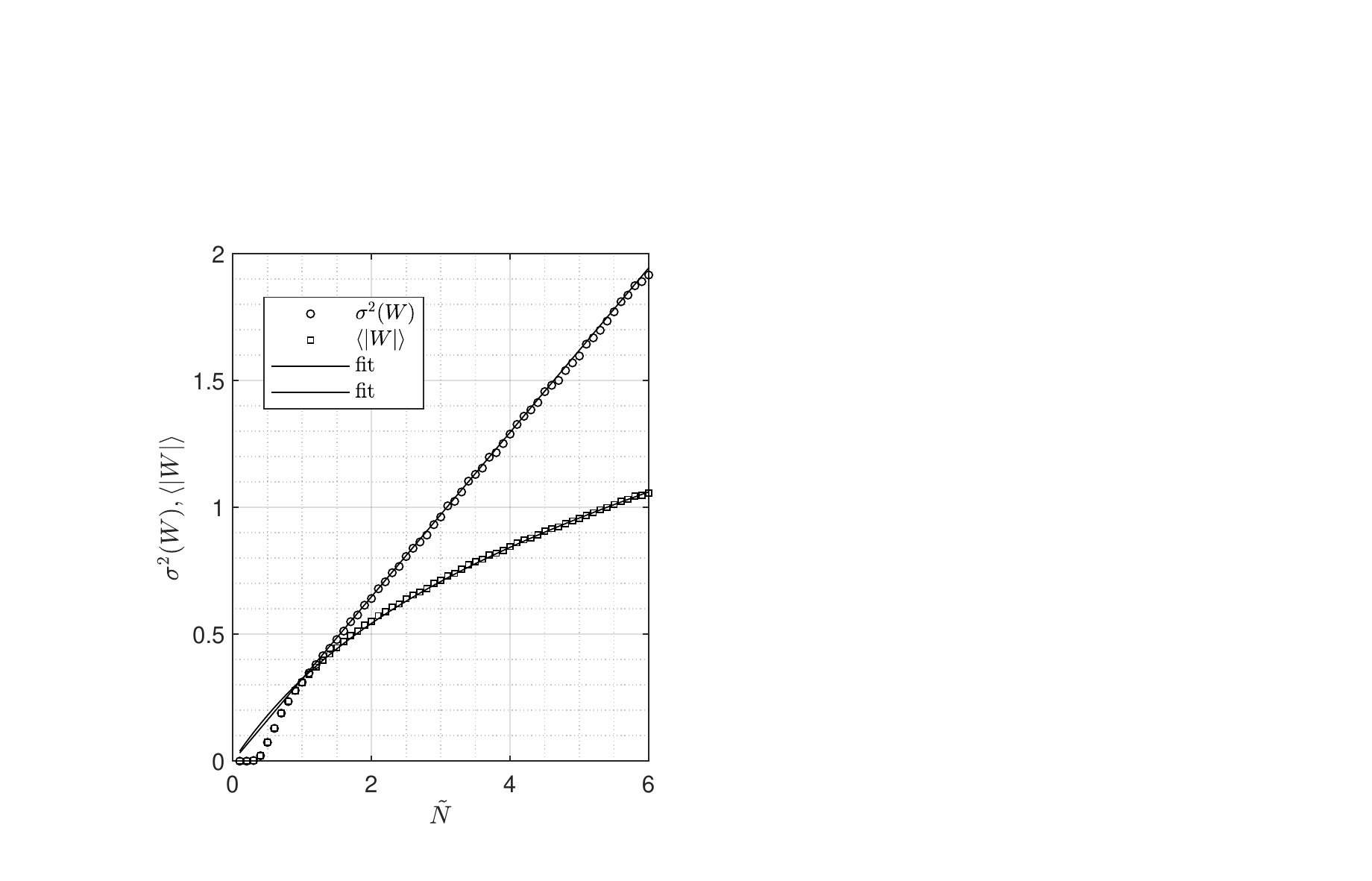}
\caption{\textbf{$\sigma^2(W)$, $\langle |W| \rangle$ and their fitted curves from trinomial distribution.} }\label{fig6}
\end{figure}





\end{bibunit}
\end{appendices}

\end{document}